\documentclass[aps,pre,floatfix,showpacs,superscriptaddress,preprint,showkeys]{revtex4}
\usepackage{latexsym}
\usepackage{amstext,amsfonts}
\usepackage{epsfig}
\usepackage{url}
\usepackage[usenames]{color}
\usepackage[hang]{footmisc}
\usepackage{slashbox}
\usepackage{multirow}
\begin{document}

\title{Jamming in complex networks with degree correlation.}

\author{Ana L. Pastore y Piontti}\affiliation{Instituto de
  Investigaciones F\'isicas de Mar del Plata (IFIMAR).Departamento de
  F\'{\i}sica, Facultad de Ciencias Exactas y Naturales, Universidad
  Nacional de Mar del Plata - CONICET, Funes 3350, 7600 Mar del Plata,
  Argentina}\author{Lidia A. Braunstein}\affiliation{Instituto de
  Investigaciones F\'isicas de Mar del Plata (IFIMAR).Departamento de
  F\'{\i}sica, Facultad de Ciencias Exactas y Naturales, Universidad
  Nacional de Mar del Plata - CONICET, Funes 3350, 7600 Mar del Plata,
  Argentina} \affiliation{Center for polymer studies, Boston
  University, Boston, MA 02215, USA}\author{Pablo
  A. Macri}\affiliation{Instituto de Investigaciones F\'isicas de Mar
  del Plata (IFIMAR).Departamento de F\'{\i}sica, Facultad de Ciencias
  Exactas y Naturales, Universidad Nacional de Mar del Plata -
  CONICET, Funes 3350, 7600 Mar del Plata, Argentina}%

\footnotetext[0]{Corresponding author: Ana L. Pastore y Piontti Tel:+54-223-475-6951\newline
\hspace{25pt}Departamento de Fisica, IFIMAR, FCEyN de la UNMDP-CONICET.\newline
\hspace{25pt}Dean Funes 3350, (7600) Mar del Plata,Argentina\newline
\hspace{25pt}fax:+54-223-475-3351\newline
\emph{E-mail address}:apastore@mdp.edu.ar (Ana L. Pastore y Piontti).}

\date{\today}

\begin{abstract}
We study the effects of the degree-degree correlations on the pressure
congestion $J$ when we apply a dynamical process on scale free complex
networks using the gradient network approach.  We find that the
pressure congestion for disassortative (assortative) networks is lower
(bigger) than the one for uncorrelated networks which allow us to
affirm that disassortative networks enhance transport through them.
This result agree with the fact that many real world transportation
networks naturally evolve to this kind of correlation. We explain our
results showing that for the disassortative case the clusters in the
gradient network turn out to be as much elongated as possible,
reducing the pressure congestion $J$ and observing the opposite
behavior for the assortative case.
Finally we apply our model to real world networks, and
the results agree with our theoretical model.

\end{abstract}

\maketitle

\section{Introduction}

It is known that a great variety of complex systems can be represented
by networks, where the nodes are the elements of the system and the
links, the interactions among them. A way to characterize a network is
trough its degree distribution. In many cases of interest
\cite{barabasi}, this distribution is often scale free, which is
characterized by a power-law degree distribution $P(k) \sim k
^{-\lambda}$, ($k \ge k_{min} $), where $k$ is the number of
connections that a node can have, $\lambda$ is the degree exponent and
$k_{min}$ is the lowest degree allowed. The degree distribution has
an important impact on the behavior of some dynamical processes taking
place on the network, specially on the congestion problem
\cite{Park,Gulbahce,NJP-us,B-PRL,PastorePRE,LaRoccaPRE,barabasi}.

Recently the attention of scientists has been focused on to another
characteristic of the complex networks: the degree
correlation.  The degree correlation can be understood as the tendency of nodes of
a certain degree to be connected with other nodes with similar
or different degree. In the first case this tendency is
called assortativity and in the last one disassortativity.

Through this property, it is possible to separate social networks from
technological networks, since the degree correlation behavior is
very different in either case. In social networks, like the physics
Co-authorship and film actors networks \cite{datasets}, nodes tend to be attached with
others of similar degree, and therefore are characterized by an
assortative degree correlation. Technological and biological networks
such as Internet and the protein-protein interaction instead have a disassortative degree
correlation.

Despite there are several models of networks proposed in the literature
that successfully reproduce many properties of real world networks,
only recently a few of them, take into account the "correlation" factor
in their construction. Newman showed that models that do not consider correlation fail to reproduce
many of the real networks properties \cite{Newman-r}.

Of particular interest is to study the effects of the degree
correlations on the dynamical processes evolving on the top of the
network. It is known that some processes, such as synchronization
\cite{Sorrentino}, transport \cite{Xue}, traffic dynamics \cite{Sun} and
growth \cite{Shao} behave differently according to the correlations present
in the substrate network where these processes spread \cite{Hu}.

From a quantitative point of view, the degree correlation can be
measured trough the neighbor connectivity \cite{Pastor-Satorras},
introducing the quantity $K_{nn}(k)=\sum_{k'}k'P(k'|k)$ where
$P(k'|k)$ is the conditional probability that an edge belonging to a
node of degree $k$ points to a node of degree $k'$. Then $K_{nn}(k)$
is the average nearest neighbor degree of a node of degree $k$. This
function increases with $k$ in the case of an assortative network,
decreases for a disassortative network and is flat for an uncorrelated
network. Other measure of the degree correlation is the Pearson
coefficient $r$ defined as

\begin{equation}
 r=\frac{M^{-1}\sum_{e}
   j_ek_e-[M^{-1}\sum_{e}\frac{1}{2}(j_e+k_e)]^2}{M^{-1}\sum_{e}\frac{1}{2}(j_e^2+k_e^2)-[M^{-1}\sum_e \frac{1}{2}
     (j_e+k_e)]^2}\;, \label{pearson-1}
\end{equation}
\\

where $j_e$, $k_e$ are the degrees of the nodes at the ends of the
$e$-th edge, with $e=1,... ,M$, between $j_e$ and $k_e$. Notice that
this expression is valid for undirected networks \cite{Newman03}.  For
assortative (disassortative) networks, $r>0$ ($r<0$), and $r=0$ for
uncorrelated networks. Even though this is an accepted quantity to
measure correlations, it should be used with caution because can hide
strong structural correlations \cite{Dorogovtsev}.  Another measure
of short range correlation is the clustering. The clustering
coefficient $c_i$, for every site $i$ gives the probability that two
nearest neighbors of node $i$ are also neighbor to each other.

Although there are many studies about degree correlations and
clustering in the literature, there is no agreement among researchers
on which topology characteristic governs a particular process that is
evolving on the network.  The problem is that the results obtained
strongly depend on the algorithms used to build the correlated
network, that in general generate clustering \cite{Newman, Miller}.

In our case, we are particularly interested in the effects of the
degree correlations in the pressure congestion of a network. To this
end we apply an algorithm that preserves the clustering and the degree
distribution $P(k)$, but allows us to change the degree correlations.
Then, through this algorithm we can isolate the effects of the degree
correlations from clustering on the pressure congestion and compare
the results with the uncorrelated case \cite{NJP-us}.
In particular we argue that real world networks of
communications evolve to a disassortative form in order to enhance the
transport trough them.

It is known \cite{NJP-us} that for the uncorrelated case the pressure
congestion increases with $\lambda$ when the process has a
relaxational component.  In \cite{NJP-us} it was shown that by
introducing a surface relaxation mechanism, congestion in SF networks
can be reduced, but the same mechanism has no effect on congestion in
the case of Erd\"{o}s Renyi random graphs \cite{ER}.

In order to study the effects of the degree correlations of the
underlying network on the transport, we measure the congestion
pressure $J$.

The congestion pressure of a network is measured in the gradient
network. The gradient direction of a node $i$ is a directed edge
pointing towards a neighbor $j$ on the substrate graph $G$, which has
the lowest value of the scalar field of its neighborhood. If $i$ has
the lowest value of $h$ in its network neighborhood, the gradient link
is a self-loop.  In other words, the gradient network is the
collection of all gradient edges on the substrate graph $G$. In the
gradient network, each node has just one outgoing link and $\ell$
incoming links. When a node has $\ell=0$, belongs to the perimeter of
a gradient network cluster. Then $J$ is the average fraction
of nodes with $\ell=0$, $J=N(\ell=0)/N$. Thus $J$ is a global
indicator of the pressure congestion and higher $J$ means more
congestion.  In Ref. \cite{NJP-us} the authors studied a dynamic
process of gradient-induced flows produced by the local gradients of a
non-degenerate scalar field ${\bf h}=h_{i=1}^N$ distributed over the
$N$ nodes belonging to $G$. They found that the dynamic process
decreases $J$ compared to the static case \cite{Znature}.  The
findings in Ref. \cite{NJP-us} were interpreted trough a structural
transition in the clusters of the corresponding gradient networks.

Recently, Pan {\it et al} \cite{Pan} studied the effects of the degree
correlations on the pressure congestion in networks without any
dynamics, using the model introduced in \cite{Gulbahce} for the
gradient network, and they found that assortative networks are less
congested than disassortative. This founding contradicts the
observation that most of the transport networks are disassortative.
Transport in real networks cannot be thought as a static process.

Even though at the present time no one has a global understanding
about what governs the evolution of complex networks, it is factual
information that many of these networks have SF degree distribution.
This include large-scale communication networks, and many biological
networks. If there is a theory to explain this, it must be based on
processes and principles that are common to all these different
systems. So it is possible to think that the network structure and hence its
evolution is tied to its main functionality,
which is transport. However, transport is ubiquitously related to
gradients, or biases distributed across the system. Therefore when a
dynamic process is applied we expect a different behavior, that could
explain why different kind of networks evolve with certain degree
correlation.

In this paper we study the effect of the degree correlations on $J$ in
SF networks when a dynamical process is applied.

\section{Theoretical Model}

In this section we discuss the construction of the substrate networks
with and without degree correlations used in this work. In order to
generate uncorrelated SF networks we implement the configurational
model \cite{ConfigModel}, with a power law degree distribution, where
the degree $k_i$ of a node $i$ is between $2=k_{min}\le k_i\le
k_{max}=\sqrt{N}$ in order to uncorrelate the original network \cite{Boguna}.
Then, to correlate the original uncorrelated network, we apply the
algorithm of rewiring links \cite{maslov,xulvi1,xulvi2}, in which, at
each step, two links of the network, connecting four different nodes
are randomly chosen.  Afterwards, the four nodes are ordered according
to their degrees. If we want to correlate the network in a assortative
mode, the links are rewired with probability $p$ in such a way that
one link connects the two nodes with larger degrees and the other link
connects the remaining nodes with smaller degrees and with probability
$1-p$ the links are rewired at random. In the opposite case, if we
want to correlate the network in a disassortative mode, with probability
$p$, one link connects the highest degree node with the lowest degree
node and the other link connects the remaining nodes. In both cases
self-loops and multiple connections are forbidden. As can be seen the
parameter $p$ controls the different degrees of assortativity or
disassortativity that a network can have. Although we cannot achieve
with this model the extremes value of $r$, to our end the values obtained are
enough demonstrative since most of real world networks correlation fall in the range of
values obtained by this model. We emphasize that this model does not
change $P(k)$ and does not
change clustering, which is very small in the original network.

\begin{figure}[htbp]
\vspace*{5mm}
\epsfig{file=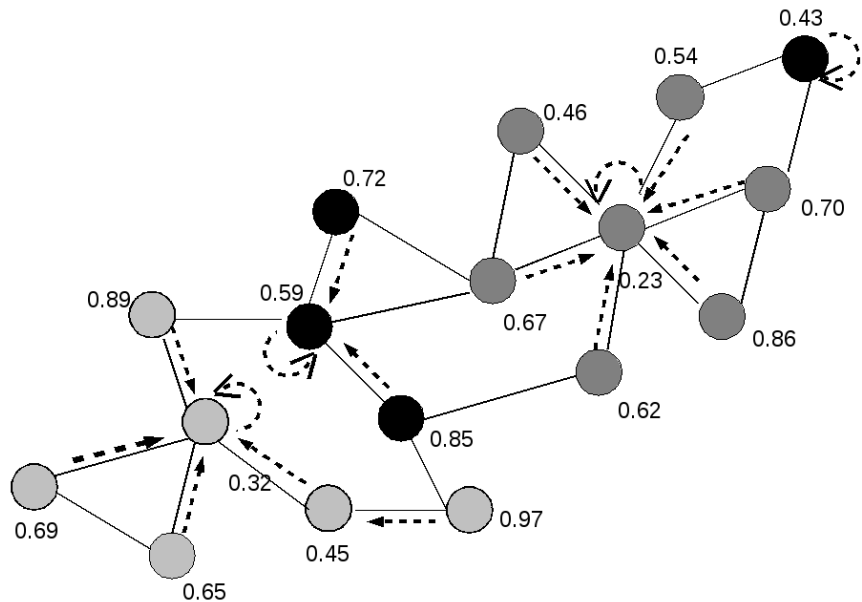,width=4.5in,angle=0} \caption{Scheme for the
  substrate network and the gradient network. The solid lines
  indicates the connections in the original network, that do not
  change during the process. The dash arrows indicate the connections
  in the gradient network and the self-loops. The numbers represent
  the scalar field of each node. Different colors represent different
  clusters in the gradient network. These clusters change as the
  dynamical process evolves. Notice that each cluster has just one
  self-loop. \label{grad}}
\end{figure}

After building the SF network, at t=0, a random scalar field $h$ is
constructed assigning to each node of the substrate network a random
scalar uniformly distributed between $0$ and $1$. In Fig. \ref{grad}
we show a scheme of the substrate network and the gradient network.
Then the scalars $h\equiv h(t)$ evolve obeying the rules of the Family
model \cite{Family,note}. This model is the simplest model of
transport due to gradients, in which at every time step a node $i$ of
the substrate is chosen at random with probability $1/N$ and it
becomes a candidate for growth. If $h_i < h_j$ for every $j$ (gradient
criterion) which is a nearest neighbor of the node $i$,
$h_i\rightarrow h_i+1$. Otherwise, if $h_i$ is not a minimum, the node
$j$ with minimum $h$ is incremented by one.  When the process reaches
the steady state of the evolution with this relaxation, we construct
the gradient network and measure $J$ \cite{PastorePRE}.

\section{Results}

\begin{figure}[htbp]
\vspace*{5mm}
\epsfig{file=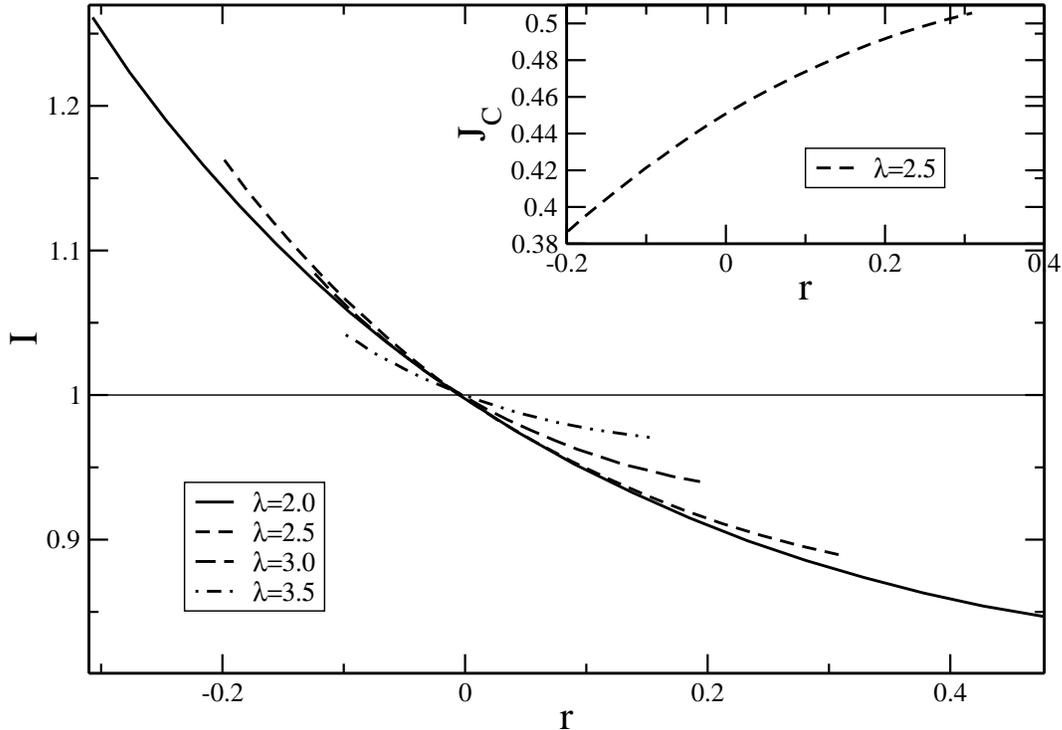,width=5.0in,angle=-90} \caption{$I=J_U/J_C$ as function
  of $r$ for different values of $\lambda$ and $N=30000$. For $r<0$
  the correlated structure has a lower jamming than the
  uncorrelated one. On the other hand, when $r>0$ the jamming is lower in the
  uncorrelated case. In the inset we plot $J_C$ as function of $r$ for
  $\lambda=2.5$. We can see that $J_C$ increases with $r$. We
  observe this behavior for different values of $\lambda$, not shown in this plot.\label{IdeR}}
\end{figure}

We run our simulations for SF networks with $k_{min}=2$ and
$N=30000$. We choose this value of $N$ in order to avoid finite size
effects on $r$ \cite{Dorogovtsev}.  We define $I=J_U/J_C$ as a
factor of improvement, where $J_U$ is the pressure congestion for
the uncorrelated network and $J_C$ for the correlated case.  Then if
$I>1$ ($I<1$) correlated networks enhance (worsen) the transport. In
Fig.~\ref{IdeR} we plot $I$ as function of $r$ for different values
of $\lambda$.  As can be seen in this figure, as $r$ decreases, $I$
increases, which means that a disassortative correlation leads, after
dynamics, to a lower value of congestion, compared to an assortative
correlation. This effect is more pronounced as $\lambda$ decreases,
implying that disassortative networks are better for transport as
$\lambda$ decreases. This result could explain the emergence of
disassortative SF networks with small $\lambda$ in communication
networks and agree with the idea that networks naturally develop
certain correlation optimizing the process that evolves on the top
of them. To give a general picture of $J_C$, in the inset of
Fig.\ref{IdeR} we plot $J_C$ as function of $r$ for $\lambda=2.5$.
We can see that $J_C$ increases with $r$. This behavior was observed
for different values of $\lambda$.

\begin{figure}[htbp]
\vspace*{5mm}
\bigskip
\epsfig{file=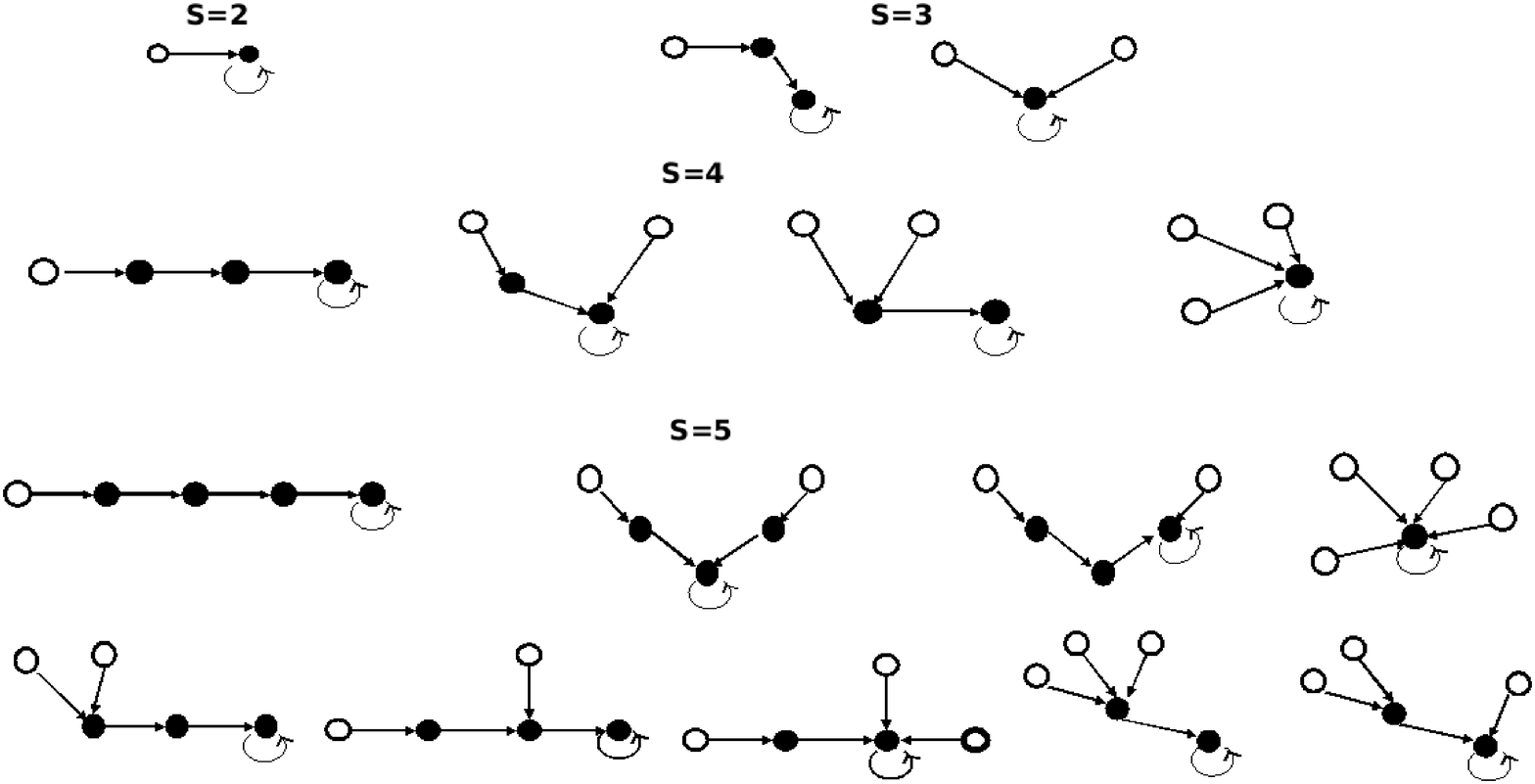,width=5.0in,angle=0}
\bigskip
\caption{Clusters configurations for different cluster size $s$ (Here
  we show just from $s=2$ to $s=5$). The white nodes are the nodes in
  the perimeter ($\ell=0$) of every cluster. As can be seen there are
  a finite number of configurations for each size $s$ and as $s$
  increases there are more possible configurations, but some of them
  have the same number of nodes in the perimeter. Notice that there is
  just one self-loop per cluster. \label{clusters}}
\end{figure}

In order to understand the previous result, we are going to consider
an ideal situation where a gradient network, in an infinite network, has every
cluster configuration equally probable. For cluster configuration we
mean every different way to connect the nodes of a cluster of size
$s$ as it is shown in Fig.~\ref{clusters} for some values of cluster
sizes.  From Fig.~\ref{clusters} it can be seen that the average
fraction of nodes in the perimeter of a cluster $\pi_s$ is almost constant as a 
function of $s$. More specifically, $\pi_s=1/2$ for $s=2,3,4$ and then decrease 
very slowly for larger sizes (for example, $\pi_7=0.482$). 
This implies that the average number of nodes in the perimeter of a gradient 
network cluster is a growing function of its size $s$. As a consequence, a 
gradient network with a large number of small clusters will have a smaller 
congestion pressure $J$ than a gradient network with a large number of big clusters.
In a real network, the number and sizes of the gradient network
clusters depends on the substrate network topology and on the relaxation
process that evolves on the top of it. 
Let's consider two archetype
cases: first an assortative network where $N$ nodes are
fully connected, {\it i.e.}, every node has $N-1$ neighbors. The
gradient network in this case is independent of the process and has
just one cluster, with one self-loop and $N-1$ nodes in the perimeter.
This example illustrate that when big clusters are present one have to expect
a large number of nodes in the perimeter and a large jamming, which in this case
is $J=N-1/N$ and tends to 1, i.e., maximum congestion. 
Second, a
disassortative network with a
star-like configuration, where one node is connected to the others $N-1$ nodes.
In this particular case there will be more or less clusters in the
gradient network depending on the efficacy of the process decongesting 
the network. For an optimal process we could obtain $N-2$ self-loops 
clusters (that means, clusters with $s=1$) and
1 cluster of size $s=2$. This example illustrate that when many small clusters 
are present one have to expect small number in the perimeter and a low jamming,
which in this case is $J=1/N$ and tends to 0, i.e., minimum congestion. Extrapolating 
the conclusions given by this two toy networks, we expect smaller clusters and lower 
jamming for a disassortative networks  comparison with an assortative one.

In the following, we are going to analyze what actually occurs with
the process and the networks implemented in this work.

\begin{figure}[htbp]
\vspace*{5mm}
\epsfig{file=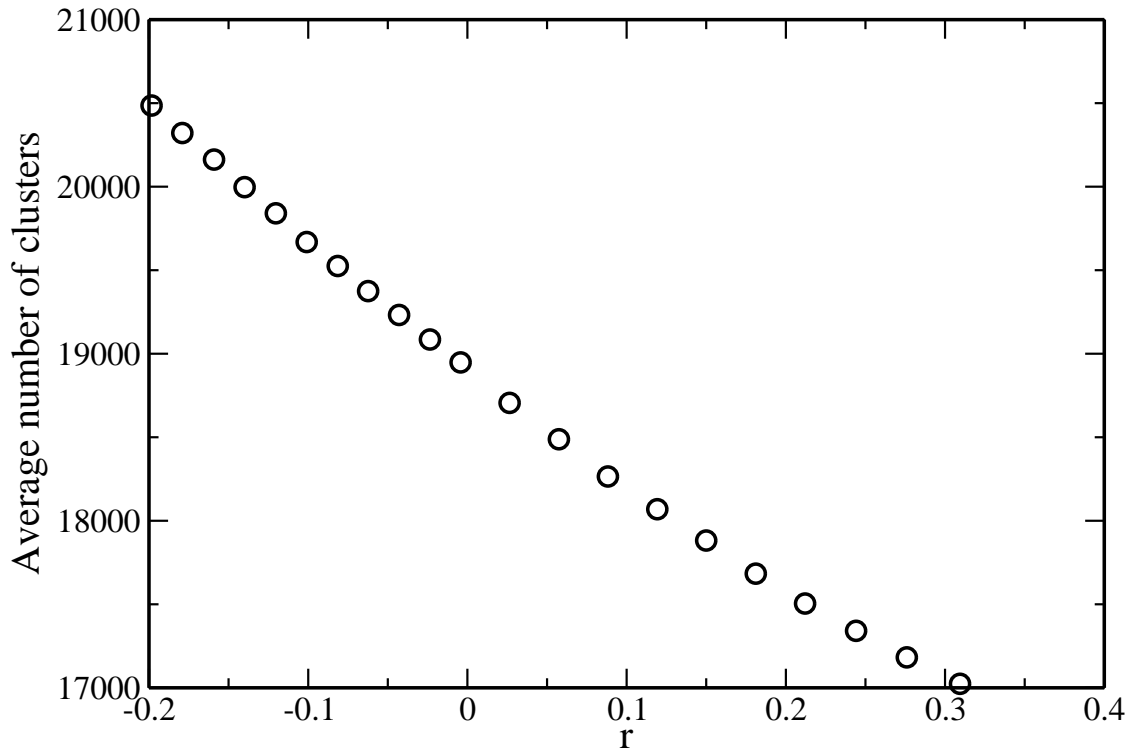,width=5.0in,angle=-90} \caption{Average number
  of clusters decreases as $r$ increases for $\lambda=2.5$ and
  $N=30000$. This result was obtained for other values of $\lambda$ as
  well, observing the same behavior. \label{dstrib}}
\end{figure}

In Fig.~\ref{dstrib} we plot the average number of clusters of the
gradient network for different values of $r$ for our model. As can be
seen, as $r$ decreases, there are more and consequently smaller 
clusters. Following our conclusions for an infinite gradient network
the presence of 
smaller clusters indicates that a low number of nodes are in the perimeter
of them and therefore it explains the lowering of the jamming observed in 
Fig. \ref{IdeR}.

\begin{figure}[htbp]
\vspace*{20mm} \epsfig{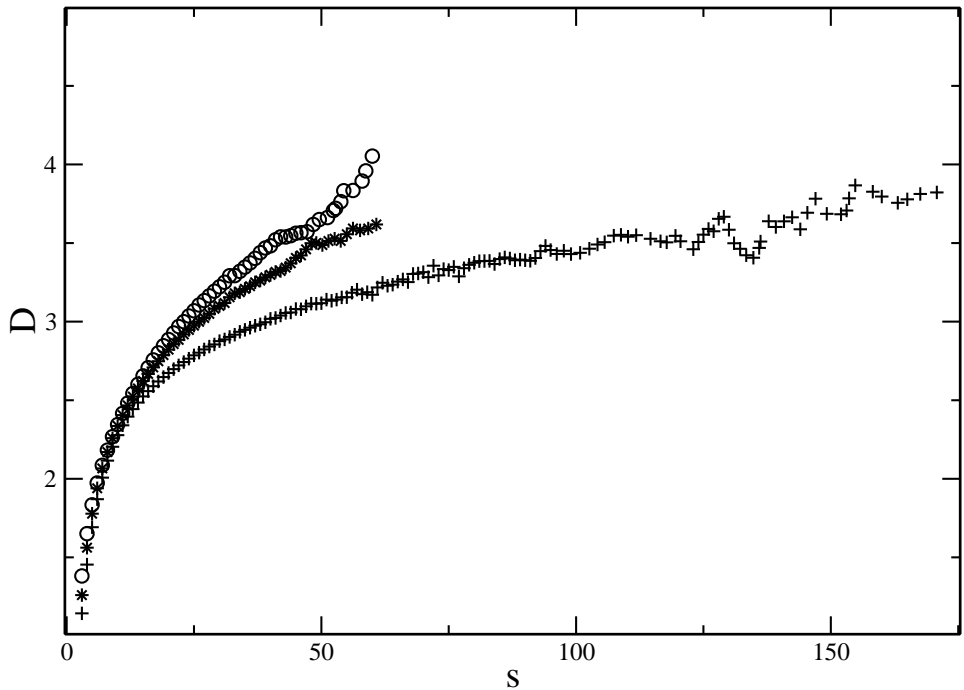} \caption{$D$
  as function of $s$ for different values of $r$: disassortative
  ($\bigcirc$ $r=-0.198$), uncorrelated ($\ast$ $r=0.004$) and
  assortative ($+$ $r=0.309$), for $\lambda=2.5$ and
  $N=30000$.  \label{diametro}}
\end{figure}

However, that conclusions were reached for infinite gradient networks
of equally probable clusters. In order to observe the effect introduced by the 
network correlation and by the relaxation process favoring some specific graphs 
of Fig. \ref{clusters} in detriment of others, in Fig. \ref{diametro} we plot
the average diameter $D(s)$ of the clusters as function of
the cluster size $s$. We define $D(s)$ as the average distance from
a perimeter node ($\ell=0$) to the self-loop of the cluster of size
$s$ (see Fig.\ref{grad}). First, we observe that assortative networks
reach a given diameter $D(s)$ for much bigger cluster sizes than disassortative
networks. But bigger clusters with the same diameter is a clear indication that
assortative networks have clusters with larger perimeters than disassortative ones,
which confirms our previous conclusions based on infinite gradient networks.
At the same time, we find that for a given cluster size $s$ dissortative networks 
show cluster with larger diameter than assortative ones. This is a new effect that
cannot be inferred from our infinite gradient network analysis and is a consequence
of the correlation in the substrate networks. The relaxation process running over 
them generate gradient network clusters that, for a given size $s$, are more elongated
in the case of disassortative networks. More elongated clusters for a given size corresponds
to a lower perimeter and hence to a lower congestion pressure $J$.

In order to see this effect in more detail, next we study the 
contribution of every cluster type (see Fig. \ref{clusters}) to the 
pressure congestion. Every cluster of size $s>1$ can have
from $1$ to $s-1$ nodes in the perimeter, and of course this result does not depend
on the correlation or the degree distribution of the network (See
Fig. \ref{clusters}). What does depend on the degree correlation is
the number of times that every configuration appears.

We observe that depending on the correlation of the substrate network,
there are certain structures favored against others: for $r<0$ there
are more self-loops clusters ($s=1$) than for $r>0$. We compute the
number of clusters with $s=1$, for the values of $r$ in Table
\ref{Table.1} before and after applying the relaxational process. We
found that, in average, before the dynamics there are $3561.07$,
$2.524$ and $1336.75$ self-loops clusters for $r=-0.198$, $r=-0.004$
and $r=0.309$ respectively. After the dynamics we find $7282.47$,
$3960.96$ and $1391.51$ self-loops for the same values of $r$. This
result means that disassortative correlations in combination with the
relaxational process contribute to the decrease in the
congestion. Something similar occurs for others values of $s$ showed
in the Table \ref{Table.1}.  Besides from Table \ref{Table.1} for any
value of $r$, for a given $s$ ($s=7$ as example) it can be seen that
the cluster configurations with $1$ and $s-1$ nodes in the perimeter
are less frequent than the others configurations. This result is due
to there are different clusters configuration which lead to the same
number of nodes in the perimeter, but for extreme cases ($1$ and
$s-1$) there is only one possible configuration.

From Table \ref{Table.1} we can also see that, after the dynamics, the pressure
congestion has its main contribution from the smaller clusters.
Computing the contribution to $J$ of the clusters from size $s=1$ to
$s=7$ and we find that the nodes in the perimeter of these clusters
represents more than the $75$\% of the total perimeter for any value of $r$ in Table
\ref{Table.1}.

The results presented in this article do not depend on the algorithm
used to build the substrate network (BA model or configurational
model) neither the dynamical process , but some effects could be due
to small features depending on the algorithm used to correlate the networks.

Finally, we want to show that our findings are reproducible in real-world networks.
Here we present results for an Internet network \cite{dataNewman} sample, the protein-protein
interaction network of yeast, and the actor movie database
\cite{datasets}. We choose these networks because they are
undirected and have a SF degree distribution: the Internet network has
$r=-0.198$ and $\lambda=2.1$, the protein interaction network has
$r=-0.156$ and $\lambda=2.4$, and the actors network has $r=0.208$ and
$\lambda=2.3$ \cite{Havlin}, so we have disassortative and
assortative networks to compare with the uncorrelated case. In order
to measure the pressure congestion in this real networks we assign a
non degenerated scalar field to each node, and then we construct the
gradient networks as it was explained previously in this paper, and perform the
relaxation process. In order to compute the improvement factor $I$, we uncorrelate the 
real networks applying the following algorithm: at each step we choose two links connecting four
different nodes and then we reconnect them at random avoiding self-loops and multiple connections.  
We found the following improvement factors $I_{Internet}=1.84\pm 0.02$ $I_{protein}=1.1\pm 0.02$ and
$I_{actors}=0.86\pm0.02$ which agree with our results found for model networks as function of the Pearson
coefficient $r$ and the power-law exponent $\alpha$.

\begin{center}
\begin{table}
\begin{tabular}{|c|c|c|c|c|c|c|c|}
\hline
 &\backslashbox{~~~~~~~~~~~$s$}{Nodes in the perimeter}&~~1~~ &~~2~~ & ~~3~~ &~ ~4~~ &~ ~5~~ &~~6~~ \\
\hline
 & \multirow{2}{0cm}{2} & 1538.29 & ×  & × & × & × & × \\
 &  & \it{2909.09} & ×  & × & × & × & × \\\cline{2-8}
\multirow{3}{2cm}{$r=-0.198$} &\multirow{2}{0cm} {3}  & 228.13 & 481.05 & × & × & × & × \\
 &  &\it{759.39} & \it{624.12} & × & × & × & × \\\cline{2-8}
&\multirow{2}{0cm} {5} & 0.65    & 38.72  & 154.35 & 45.39  & ×     & × \\
  &  & \it{13.12} & \it{195.73} & \it{237.28} & \it{1.17} & × & × \\ \cline{2-8}
  &\multirow{2}{0cm} {7}  & 0.01    & 0.34   & 8.8  & 46.39 & 57.03 & 8.01 \\
 &  & \it{0.13} & \it{8.75} & \it{60.06} & \it{87.93} & \it{31.07} & \it{0.74} \\
\hline
\hline
&\multirow{2}{0cm} {2} & 1979.30 & ×  & × & × & × & × \\
 &  &\it{2925.49} & ×  & × & × & × & × \\\cline{2-8}
\multirow{3}{2cm}{$r=-0.004$}   &\multirow{2}{0cm} {3} & 286.29  & 829.57 & × & × & × & × \\
&  & \it{666.95} & \it{961.39} & × & × & × & × \\ \cline{2-8}
 & \multirow{2}{0cm}{5} & 1.66    & 71.69  & 188.85 & 48.1  & ×     & × \\
 & & \it{14.39} & \it{249.5} & \it{250.96} & \it{20.11} & × & × \\ \cline{2-8}
& \multirow{2}{0cm}{7} & 0.01 & 1.08 & 17.13 & 58.82 & 50.99 & 6.41\\
  & & \it{0.14} & \it{13.64} & \it{79.80} & \it{97.14} & \it{28.28} & \it{0.8} \\
\hline
\hline
& \multirow{2}{0cm}{2} & 1879.64 & ×  & × & × & × & × \\
&  & \it{1932.18} & ×  & × & × & × & × \\ \cline{2-8}
\multirow{3}{2cm}{$r=0.309$}  &\multirow{2}{0cm} {3} & 326.62 & 1443.89 & × & × & × & × \\
&  & \it{426.59} & \it{1263.94} & × & × & × & × \\ \cline{2-8}
& \multirow{2}{0cm}{5} & 4.29 & 170.1 & 271.22 & 59.48 & × & × \\
&  & \it{12.16} & \it{316.10} & \it{260.68} & \it{32.30} & × & × \\ \cline{2-8}
 &\multirow{2}{0cm}{7} & 0.01 & 5.94 & 44.32 & 83.14 & 81.43 & 4.45\\
 &  & \it{0.17} & \it{27.41} & \it{97.84} & \it{94.68} & \it{27.94} & \it{1.91}\\
\hline
\end{tabular}
\caption{For different values of $r$ we compute the number of every
  possible configuration, before and after (in italics) applying the dynamics, for
  different values of $s$ for $N=30000$ and $\lambda=2.5$. In this
  table we are only showing some values of $s$.\label{Table.1}}
\end{table}
\end{center}

\section{Conclusions}

In the present work we studied the effects of the degree correlations to
the congestion on SF complex networks when a dynamic process is
applied.  As a result we found that disassortative networks are better
for transport compared to uncorrelated networks. This
result could explain why real world networks of transport have $r<0$.

We also showed that the same relaxational dynamics has a bigger effect
reducing the congestion in networks with lower values of $\lambda$.
This result agree with the fact that real transportation networks
evolve to structures with $2<\lambda<3$ and $r<0$.

We explained our results showing that for $r<0$ the clusters in the
gradient network turn out to be as much elongated as possible,
reducing the perimeter and hence the pressure congestion $J$ and
observing the opposite behavior for $r>0$. We showed this computing
the times that every cluster configuration  appeared for some values
of $s$.

Finally we applied our model to some real networks and the results
show that these networks evolve to topologies that optimize certain
processes.


\begin{thebibliography}{99}
\bibitem{note} It is common to map this synchronization model into an interface growth model.
\end{thebibliography}
\end{document}